\documentclass[aps,prb,twocolumn]{revtex4}
\usepackage{amssymb}
\usepackage{graphicx}

\begin{document}
\title{ First-order magneto-structural transition in single crystals of the honeycomb lattice Ruthenate Li$_{2}$RuO$_3$}
\author{Kavita Mehlawat and Yogesh Singh}
\affiliation{Indian Institute of Science Education and Research (IISER) Mohali, Knowledge city, Sector 81, Mohali 140306, India}

\date{\today}

\begin{abstract}
Li$_2$RuO$_3$ is known to crystallize in either $C2/m$ or $P2_1/m$ structures at room temperature.  We report the first single crystal growth of Li$_2$RuO$_3$ and Na substituted crystals (Li$_{0.95}$Na$_{0.05}$)$_{2}$RuO$_3$ crystallizing in the $P2_1/m$ structure where a magneto-structural transition is observed at high temperatures.  Using high temperature ($T \leq 1000$~K) magnetic susceptibility $\chi$ measurements we study the magnetic anisotropy across the magneto-structural transition.  Our results show for the first time that for Li$_2$RuO$_3$ the magnetic and structural transitions most likely occur at slightly different temperatures.  The structural transition which is first order-like occurs first ($T \approx 570$~K) and drives the magnetic transition ($T \approx 540$~K).  Rather surprisingly, just $5\%$ Na substitution for Li affects the magneto-structural transition in an interesting way.  The first order transition temperature stays $\approx 540$~K, the magnetic anisotropy is reversed, and the Ru-Ru dimerization pattern changes from two short and four long Ru-Ru bonds per honeycomb in an armchair pattern for Li$_2$RuO$_3$ to four short and two long bonds per honeycomb in (Li$_{0.95}$Na$_{0.05}$)$_{2}$RuO$_3$ which can be viewed as two inter-penetrating armchair patterns. 

\end{abstract}

\maketitle
\section{Introduction}
\label{sec:INTRO}
Mott insulators with spin-orbit (SO) coupling have recently been topics of great interest because of the plethora of novel phases and behaviors they are expected to exhibit \cite{Jackeli2009, Shitade2009, Pesin2010}.  Iridium based transition metal oxides are ideal systems to investigate the novel behaviors predicted to arise due to the interplay of electron correlations and SO coupling \cite{Pesin2010,Wan2011}.  In recent years honeycomb lattice iridates  A$_{2}$IrO$ _{3} $ (A = Na,Li) have been subjects of intense scrutiny which was fuelled initially by the suggestion of exotic topological properties and Quantum Spin Hall effect ~\cite{Kitaev2006, Shitade2009} and by suggestions that these could be realizations of the Kitaev-Heisenberg model  \cite{Jackeli2009, Chaloupka2010}.  Na$_2$IrO$_3$ was found to undergo novel magnetic ordering at low temperatures suggesting that it wasn't situated in the strong Kitaev limit where a spin liquid was expected \cite{Singh2010, Singh2012}.  Recently however, evidence for dominant bond-directional magnetic exchange and real space-magnetic moment locking has been found in Na$_2$IrO$_3$ \cite{Chun2015}.  For ruthenates, the spin-orbit coupling is expected to be comparitively smaller.  Nevertheless the compound $\alpha$--RuCl$_3$, which has a network of Ru$^{3+}$ $S = 1/2$ moments on a honeycomb lattice has recently been studied in the quest for a candidate Kitaev material \cite{Banerjee2016}.  Observations of a quasi-continuum of excitations in Raman scattering for both Na$_2$IrO$_3$ and $\alpha$-RuCl$_3$ has been argued to be evidence for proximity to the quantum spin liquid state in the dominant Kitaev limit \cite{Gupta2014, Sandilands2015}.  More recently, when Ir$^{4+}$ was partially replaced by Ru$^{4+}$ in A$_{2}$Ir$_{1-x}$Ru$_{x}$O$ _{3} $ (A = Na,Li), the materials were found to remain insulating and a spin-glass state is observed at low temperatures highlighting the presence of competing interactions and phases in the parent iridate compounds \cite{Mehlawat2015,Lei2014}.  

The ruthenate family A$_{2}$RuO$ _{3}$ (A = Na,Li) is also known to adopt a honeycomb lattice structure but with nominal $S = 1$ moments arising from the low-spin state of Ru$ ^{4+}$. Polycrystalline samples of Na$_2$RuO$_3$ were reported to crystallize in the $C2/c$ structure \cite{Mogare2004} similar to early reports on Na$_2$IrO$_3$ \cite{Singh2010}.  More recently, single crystals of Na$_2$RuO$_3$ were synthesized and found to crystallize in the related but more symmetric $C2/m$ structure \cite{Wang2014}.  Single crystal Na$_2$RuO$_3$ was found to be a local moment magnet which orders antiferromagnetically below $T_N = 30$~K \cite{Wang2014}.   

The structure and magnetic properties of Li$_2$RuO$_3$ seem to be very sensitive to synthesis conditions and  quality of samples.  Initial reports on polycrystalline samples suggested a room temperature $C2/c$ monoclinic structure and metallic paramagnetic behavior below $T = 300$~K \cite{Dulac1970, James1988, Felner2002}.  Later a comprehensive study on polycrystalline samples of Li$_{2}$RuO$ _{3}$ revealed an unusual second order structural phase transition near $T \approx 540$~K from a nearly perfect honeycomb lattice $C2/m$ structure at high temperature to a low temperature structure with a distorted honeycomb lattice P2$ _{1}/m$ \cite{Miura2007}.  This structural transition was acompanied by an increase in resistance and loss of magnetization.  Nearly perfect hexagons of the high temperature $C2/m$ phase undergo strong distortion, leading to a low temperature structure with significant shortening of one of the three inequivalent Ru-Ru bonds on each honeycomb \cite{Miura2007}.  Based on DFT calculations on the low and high temperature structures it was proposed that Li$_{2}$RuO$ _{3}$ undergoes a transition from a highly correlated metal to a molecular orbital insulator involving Ru-Ru dimerization and spin-singlet formation \cite{Miura2007, Miura2009}.  An alternative mechanism of spin-singlet formation driven by magnetoelastic coupling has also been proposed \cite{Jackeli2008}.  The evolution of the structural Ru-Ru dimers across the phase transition has been studied recently using pair distribution function (PDF) analysis of high energy powder X-ray data.  The PDF analysis allows the tracking of short-ranged structural order.  It was found that dynamically fluctuating dimers survive at temperatures well above the transition temperature $T \approx 540$~K \cite{Kimber2014}.  This suggests a scenario where a valence bond crystal in the low temperature phase melts into a valence bond liquid at high temperatures. Such a scenario is supported by recent Ru site dilution experiments \cite{Park2016}.  An electronic structure study has highlighted the importance of electronic correlations and proposed that a combination of local-moment behavior and molecular orbital formation could be the correct picture for this material \cite{Pchelkina2015}

Recently a careful study of the effect of synthesis conditions on the structure and magnetic behavior of polycrysatalline samples has been carried out \cite{Segura2016}.  It was found that all samples crystallized in the $P2_1/m$ structure at room temperature and showed the Ru-dimerization transition at high temperatures.  However, the details of the structure and the magnetic properties strongly depends on the synthesis conditions.  The best quality samples revealed that the magneto-structural transition is first-order in nature with a much higher onset temperature of $\approx 550$~K \cite{Segura2016}.

Lastly, single crystals of Li$_2$RuO$_3$ have recently been synthesized.  The crystals are found to crystallize at room temperature in either the $C2/m$ or the $P2_1/m$ structures depending on synthesis conditions.  However, in complete contrast to all existing polycrystalline work \cite{Miura2007, Kimber2014, Park2016, Segura2016}, neither of these crystals show the magneto-structural transitions at high temperature.  They instead show Curie-Weiss behavior below $300$~K and magnetic ordering at low temperatures into supposedly antiferromagnetic states \cite{Wang2014}.

In this work we report the first crystal growth of Li$_2$RuO$_3$ and $5\%$ Na substituted Li$_2$RuO$_3$ crystallizing in the $P2_1/m$ structure at room temperature and showing the magneto-structural transition at high temperatures.  We are therefore able to study for the first time the magnetic anisotropy across the high temperature magneto-structural transition. We observe that for Li$_2$RuO$_3$ the transition might occur in two steps with a first-order structural transition occurring first (onset $\approx 570$~K) which then drives the magnetic Ru-Ru dimerization transition ($\approx 540$~K).  Replacing just $5\%$ Li by Na leads to a reversal of the magnetic anisotropy although the first-order magneto-structural transition is still seen at $\approx 540$~K\@.  Room temperature structural studies show that the Ru-Ru structural dimerization arrangement is also changed in the Na substituted samples.  While the Li$_2$RuO$_3$ shows 2 short and 4 long Ru-Ru bonds on each honeycomb in an armchair pattern as previously seen \cite{Miura2007,Jackeli2008}, the Na doped samples show 4 short and 2 long Ru-Ru bonds on each honeycomb in an arrangement which can be viewed as two inter-penetrating armchair patterns.

\section{EXPERIMENTAL DETAILS}
\label{sec:EXPT}
The single crystalline samples of (Li$_{1-x}$Na$_{x}$)$ _{2} $RuO$_3$  ($x = 0, 0.05$) have been synthesized.  The starting materials were Li$_2$CO$_3$ (99.995$\%$ Alfa Aesar, Na$_2$CO$_3$ (99.995$\%$ Alfa Aesar) and Ru metal powder (99.95$\%$ Alfa Aesar). Single crystals were grown using a self flux growth method.  Off-stoichiometric amounts of starting materials were mixed and placed in an alumina crucible with a lid, heated to $750~^\circ$C for 24~h for calcination and then furnace cooled to room temperature. Crystal growth was done by keeping the calcined mixture for long periods ($70$--$80$~h) at temperatures between $1000~^\circ$C -- $1050~^\circ$C after which the furnace is turned off and allowed to cool to room temperature. Shiny plate like single crystals (size $\sim 0.5 \times 0.5 \times 0.03$) were found to grow on top of semi-melted polycrystalline powder.  Growth of crystals with higher Na concentrations were tried but were not successful.  The structure and composition of the resulting samples were checked by single-crystal and powder x-ray diffraction (PXRD), and chemical analysis using energy dispersive x-ray (EDX) analysis with a JEOL scanning electron microscope (SEM).  The PXRD was obtained by a Rigaku diffractometer with Cu K$_{\alpha}$ radiation in 2$ \theta $ range from 10$^\circ$ to 90$^\circ$ with 0.02$^\circ$ step size. Anisotropic magnetic susceptibility measurements upto $T = 1000$~K were measured on a collection of co-aligned crystals with total mass $\approx 12$~mg using the VSM Oven option on a Quantum Design physical property measurement system. 

\section{RESULTS}

\begin{figure}[t]   
\includegraphics[width= 3 in]{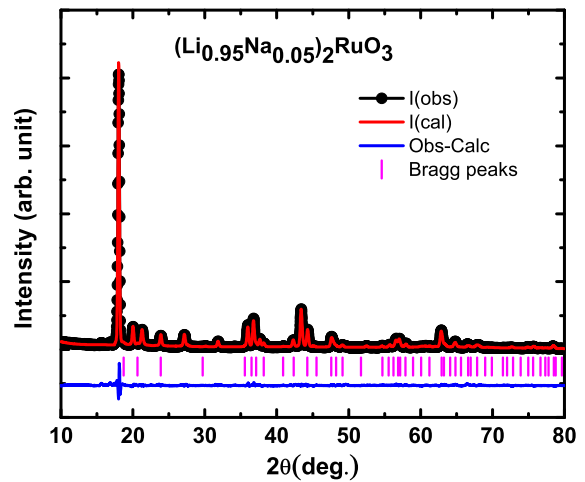}   
\caption{(Color online) Rietveld refinements of powder x-ray diffraction data for (Li$_{0.95}$Na$_{0.05}$)$ _{2} $RuO$_3$. The solid circles represent the observed data, the solid lines through the data represent the fitted pattern, the vertical bars represent the peak positions, and the solid curve below the vertical bars is the difference between the observed and the fitted patterns. 
\label{Fig-1}}
\end{figure}

\subsection{Crystal Structure and Chemical Analysis}
From room temperature single crystal and powder x-ray diffraction, we conclude that all samples adopt the $P2 _{1} $/m space group.  A full single crystal refinement was not possible because the crystals have multiple twins rotated around the $c^*$ axis.  However, it was possible to determine the space group and cell parameters using single crystal diffraction.  Cell parameters were also obtained by performing Rietveld refinements of the PXRD patterns obtained on the crushed crystals.  Fig.\ref{Fig-1} shows representative results of Rietveld refinement of the PXRD patterns for (Li$_{0.95}$Na$_{0.05}$)$ _{2} $RuO$_3$.  The fractional atomic positions obtained from the refinement are given in Table~\ref{Table-Structural-parameters}.  The unit cell parameters and the relevant bond lengths extracted from Rietveld refinement of the powder diffraction data are listed in Table~\ref{Table-lattice-parameters}.  The cell parameters change significantly (specially the $a$-axis) as Li is partially replaced by Na.  The presence of Na in the doped crystals and its concentration relative to Ru was confirmed using energy dispersive X-ray spectroscopy on several spots on the same crystal and on several crystals and was found to be close to the nominal concentration targeted in the starting material. 

\begin{figure}[t]   
\includegraphics[width= 3 in]{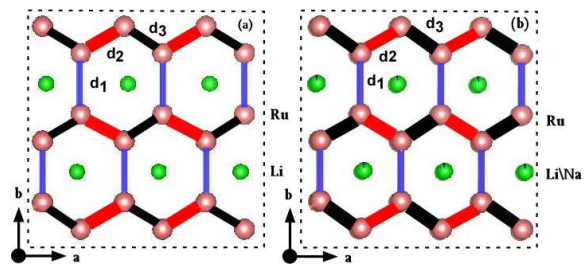}   
\caption{(Color online) Room temperature structure of (a) Li$_2$RuO$_3$ and (b) (Li$_{0.95}$Na$_{0.05}$)$ _{2} $RuO$_3$ viewed perpendicular to the Ru honeycomb network in the $ab$-plane.  There are three inequivalent Ru-Ru bonds in the honeycomb network labled as $d_1$ (blue), $d_2$ (red), and $d_3$ (black).  For Li$_2$RuO$_3$, the Ru-Ru dimerization happens on the $d_2$ bonds (shown as the thicker red bonds in (a)) which are considerably shorter than $d_1$ and $d_3$ which are of similar length. The armchair pattern observed for Li$_2$RuO$_3$ is consistent with that observed earlier \cite{Miura2007}. For (Li$_{0.95}$Na$_{0.05}$)$ _{2} $RuO$_3$ the dimerization pattern changes and there are two short bonds $d_2$ (thick red) and $d_3$ (thick black) and one long bond $d_1$ (thin blue).  The dimer arrangement can be viewed as two inter-penetrating armchair patterns.   
\label{Fig-structure}}
\end{figure}

The room temperature crystal structure of Li$_2$RuO$_3$ and Li viewed perpendicular to the honeycomb planes is shown in Fig.~\ref{Fig-structure} to highlight the Ru-Ru dimerization pattern.  For Li$_2$RuO$_3$ we find, consistent with previous work, that one ($d_2$) out of the three inequivalent Ru-Ru bonds is considerably shorter compared to the other two which are of similar lengths.  Surprisingly, for just 5\% Na substitution for Li, the dimerization pattern changes and we now have two short ($d_2$ and $d_3$) and one long bonds.  The Ru-Ru bond lengths are given in Table~\ref{Table-lattice-parameters} and the dimerization patterns shown in Figs.~\ref{Fig-structure}~(a) and ~(b), respectively.  For Li$_2$RuO$_3$ as observed \cite{Miura2007} and explained \cite{Jackeli2008} previously the dimers on the $d_2$ bond form an armchair pattern.  For the Na substituted sample, both $d_2$ and $d_3$ bonds dimerize and form inter-penetrating armchairs which run along the $a$-axis.     

\begin{table}
\caption{Wyckoff position for (Li$_{0.95}$Na$_{0.05}$)$ _{2} $RuO$_3$  obtained from Rietveld refinements of polycrystal x-ray data at 300 K}.
\begin{ruledtabular}
\begin{tabular}{|ccccccc|}
$Atom$ & $Wyckoff$ & $x$  & $y$ & $z $& $Occ$ &   $B$ $({\AA})$  \\ \hline  
Ru& 4f & 0.2467(7) & 0.0776(8) & -0.0038(7) & 1 & 0.0265 \\
Li1 & 2e & 0.7857(5) & 0.25 & -0.0295(8) & 0.95 & 0.0800 \\
Na1 &  2e  & 0.7857(5) & 0.25 & -0.0295(8) & 0.05 & 0.0900  \\
Li2 &  4f & 0.0661(3) & 0.25 & 0.6213(7) & 1 & 0.0034 \\
Li3 & 2e & 0.6887(3) & 0.0523(5) & 0.4685(6) & 1 & 0.0020 \\
O1 & 4f & 0.7812(6) & 0.0644(7) & 0.2831(8) & 1 & 0.0043 \\
O2 & 4f & 0.7502(5) & 0.0957(7) & 0.7931(2) & 1  & 0.0060 \\
O3 &  2e & 0.3124(7) & 0.25 & 0.2688(5) & 1 & 0.0088 \\
O4 & 2e & 0.2396(8) & 0.25 & 0.2373(4) & 1 & 0.0080 \\
\end{tabular}
\end{ruledtabular}
\label{Table-Structural-parameters}
\end{table}

\begin{table}
\caption{Summary of Lattice Parameters and relevant bond lengths of
(Li$_{1-x}$Na$_{x}$)$ _{2} $RuO$_3$ ($x \approx 0,
0.05 $)}
\begin{ruledtabular}
\begin{tabular}{|c|c|c|}
Sample & Li$ _{2} $RuO$_3$ & (Li$_{0.95}$Na$_{0.05}$)$ _{2} $RuO$_3$ \\ \hline
$Space $ $Group$  & P2$ _{1} $ /m  & P2$ _{1} $ /m   \\ \hline
$a $ $({\AA}) $ &  4.920(4) & 4.934(5)   \\ \hline
$b $ $({\AA})$ & 8.781(7) & 8.774(4)    \\ \hline
$c $ $({\AA})$ & 5.893(3) & 5.895(6)    \\ \hline
$\beta $ $(deg)$  &  124.36(4)  & 124.42(6)   \\ \hline
$V$  $({\AA})$    & 210.452( 5)& 210.452(5)    \\ \hline
$Ru-Ru$  $({\AA})$ &  &  \\ \hline
d$_{1}$   & 3.024 & 3.025  \\ \hline
d$_{2}$   & 2.632 & 2.812    \\ \hline
d$_{3}$   & 2.937 & 2.823    \\ 
\end{tabular}
\end{ruledtabular}
\label{Table-lattice-parameters}
\end{table}

\begin{figure}[t]   
\includegraphics[width= 3 in]{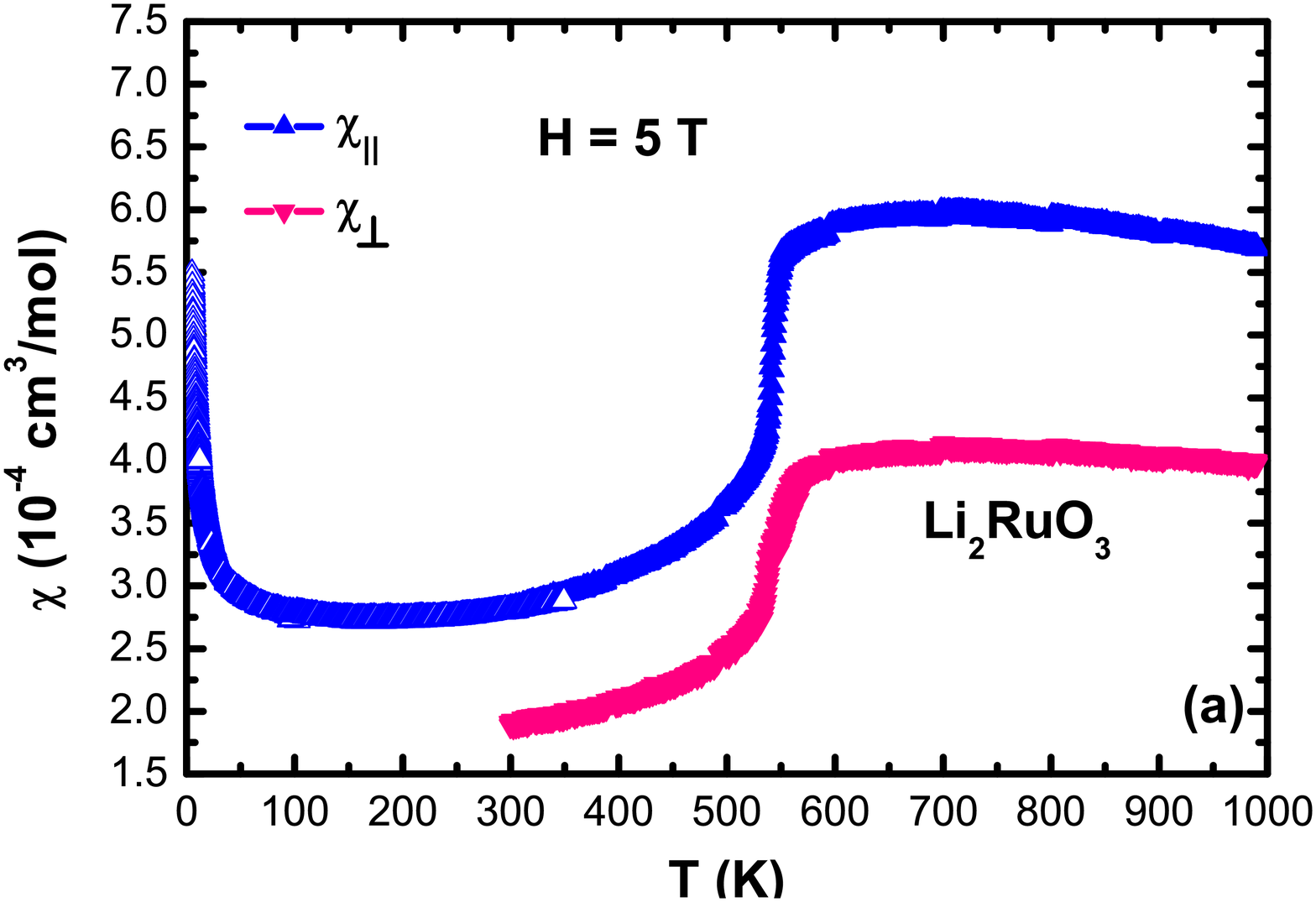}
\includegraphics[width= 3 in]{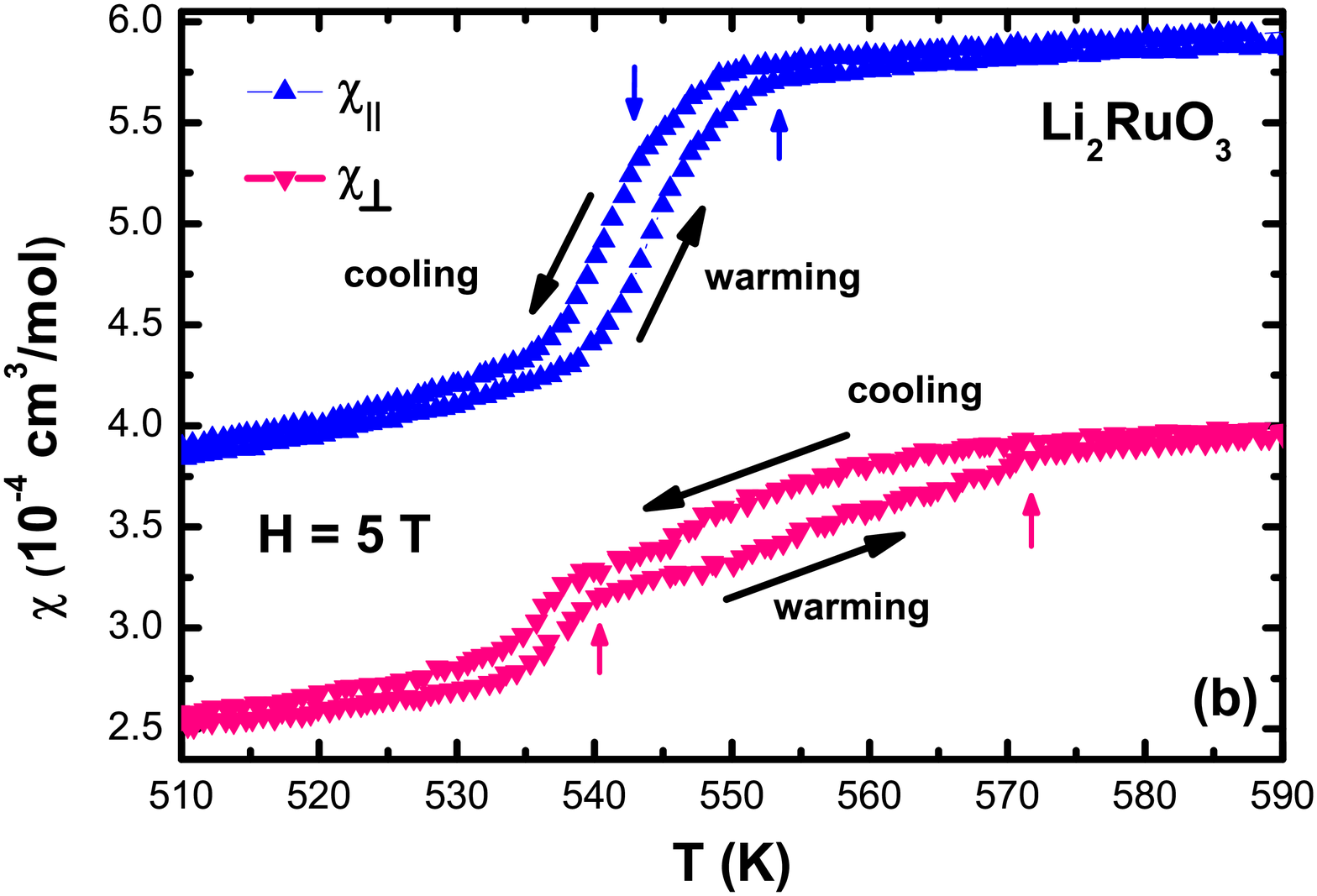}      
\caption{(Color online)  (a) Anisotropic Magnetic susceptibility $\chi_{||}$ and $\chi_{\perp}$ versus $T$ measured at in a magnetic field of $5$~T for Li$_2$RuO$_3$ between $T = 2$~K and 1000~K\@. (b) $\chi_{||}$ and $\chi_{\perp}$ versus $T$ in the temperature range $510$~K to $590$~K to highlight the behaviour near the transition. 
\label{Fig-2}}
\end{figure}

\begin{figure}[t]   
\includegraphics[width= 3 in]{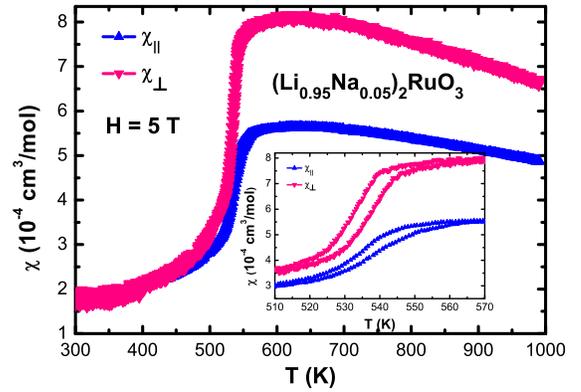}      
\caption{(Color online)  Anisotropic Magnetic susceptibility $\chi_{||}$ and $\chi_{\perp}$ versus $T$ measured at in a magnetic field of $5$~T for (Li$_{0.95}$Na$_{0.05}$)$ _{2} $RuO$_3$ between $T = 300$~K and 1000~K\@. The inset shows the $\chi_{||}$ and $\chi_{\perp}$ versus $T$ in the temperature range $510$~K to $570$~K to highlight the behaviour near the transition.  
\label{Fig-3}}
\end{figure}

\subsection{ Magnetic susceptibility}
\subsubsection{Li$_2$RuO$_3$}
The magnetic susceptibility $\chi$ versus T data for Li$ _{2} $RuO$_3$ measured in an applied magnetic field $H=5$~T applied parallel to the honeycomb plane ($\chi_{||}$) or perpendicular to the honeycomb plane ($\chi_{\perp}$) are shown in Fig.~\ref{Fig-2}.  Figure~\ref{Fig-2}(a) shows the $\chi_{\perp}$ data from $300$~K to $1000$~K and the $\chi_{||}$ data from $2$~K to $1000$~K\@.  Both sets of data were measured while cooling down from $1000$~K\@.  The first thing to note is that $\chi_{||}>\chi_{\perp}$ for all temperatures.  The $\chi(T)$ behavior at high temperatures is not Curie-Weiss like as expected for a paramagnet.  Instead the $\chi(T)$ behavior is consistent with a quasi-two-dimensional magnetic system having stronger in-plane interactions.  We also see evidence for a transition involving an abrupt drop in $\chi$ below about $550$~K\@.  This is a signature of the magneto-structural transition observed previously for polycrystalline samples \cite{Miura2007}.  The magneto-structural transition has been previously reported to involve a structural change below $540$~K from $C2/m$ to $P2_1/m$ symmetry and a simultaneous Ru-Ru dimerizations with spin-singlet formation \cite{Miura2007}.  The abrupt drop in $\chi$ at the transition is consistent with Ru-Ru spin-singlet formation.  The magnitude of the drop can be quantified by $\chi_{min}/\chi_{max}$ and is $\approx 0.45$ for both $\chi_{||}$ and $\chi_{\perp}$.  Below $300$~K, the $\chi(T)$ is $T$ independent and small but finite.  This $T$ independent finite value ($\chi_{||} \approx 2.75 \times 10^{-4}$~cm$^3$/mol) is most likely a Van Vleck paramagnetic contribution \cite{Khaliullin2013}.  

In Fig.~\ref{Fig-2}~(b) we show the $\chi_{\perp}$ and $\chi_{||}$ data on an expanded scale around the region of the transition.  Data were recorded while warming from $300$~K to $1000$~K and then while cooling back again at a rate of $5$~K/min.  We see that there is a thermal hysteresis between the warming and cooling data indicating the first order nature of the phase transition.  The transition temperatures obtained by taking derivatives of the data (not shown) are listed in Table~\ref{Table-Transition Temperature}.  For $\chi_{||}$ we get the transition temperatures $544$~K for warming and maybe a double transition at $540$~K and $546$~K for cooling measurements.  The thermal hysteresis is about $5$~K\@.  For $\chi_{\perp}$ the situation is more complex.  The transition clearly happens in two steps as indicated by the vertical arrows close to the data in Fig.~\ref{Fig-2}(b) signalling the onset of the two transitions.  A derivative of the $\chi_{\perp}$ data shows two peaks which are taken as the approximate transition temperatures and listed in Table~\ref{Table-Transition Temperature}.  We note that the lower transition is sharp and is not accompanied by any significant thermal hysteresis whereas the higher temperature transition is broad and clearly hysteretic.  The hysteresis in the higher temperature transition is about $6$~K as observed for the $\chi_{||}$ data.  We will return to a discussion of these data in a later section.  
 
\begin{table}

\caption{ Temperatures of the peaks in $d\chi/dT$ for $\chi_{||}$ and $\chi_{\perp}$ of single crystalline Li$_{2} $RuO$_3$ at $H = 5$~T}

\begin{ruledtabular}

\begin{tabular}{|ccc|}
Magnetic susceptibilities & $T_{1}$ & $T_{2}$   \\ \hline  
Li$_2$RuO$_3$ \\ \hline
$\chi_{||}$ (heating) & $540$~K &   $546$~K \\
$\chi_{||}$ (cooling) & &   544$K$  \\
$\chi_{\perp}$ (heating) & $538.3$~K  & $561$~K    \\
$\chi_{\perp}$ (cooling) & $537.8$~K  & $555$~K    \\ \hline
(Li$_{0.095}$Na$_{0.05}$)$_2$RuO$_3$  \\ \hline  
$\chi_{||}$ (heating) & $542$~K &    \\
$\chi_{||}$ (cooling) & $535$~K & \\
$\chi_{\perp}$ (heating) & $538$~K  &    \\
$\chi_{\perp}$ (cooling) & $531$~K  &    \\  
 
\end{tabular}

\end{ruledtabular}
\label{Table-Transition Temperature}
\end{table}

\subsubsection{(Li$_{0.95}$Na$_{0.05}$)$ _{2} $RuO$_3$}  
The magnetic susceptibility $\chi$ versus T data for (Li$_{0.95}$Na$_{0.05}$)$ _{2}$RuO$_3$ measured between $300$~K to $1000$~K in an applied magnetic field $H=5$~T applied parallel to the honeycomb plane ($\chi_{||}$) or perpendicular to the honeycomb plane ($\chi_{\perp}$) are shown in the main panel in Fig.~\ref{Fig-3}.  Surprisingly, with only a $5\%$ Na substitution for Li, the anisotropy is reversed ($\chi_{\perp} > \chi_{||}$) compared to what was observed for Li$_2$RuO$_3$.  The magneto-structural transition can be seen in both sets of data.  The Fig.~\ref{Fig-3}~inset shows the $\chi_{||}$ and $\chi_{\perp}$ data in the temperature range $510$~K to $570$~K to highlight the transition.  Data were recorded while warming from $300$~K to $1000$~K and then while cooling back again at a rate of $5$~K/min.  We see that there is a thermal hysteresis between the warming and cooling data indicating that the first order nature of the phase transition persists in Na substituted samples.  Peaks in the derivatives of the $\chi(T)$ data are taken as the approximate transition temperatures and are given in Table~\ref{Table-Transition Temperature}.    

\section{SUMMARY AND DISCUSSION}
We have grown the first single crystals of (Li$_{1-x}$Na$_{x}$)$ _{2} $RuO$_3$ ($x = 0, 0.05$) crystallizing in the $P2_1/m$ structure at room temperature and showing a magneto-structural transition at high temperatures.  Using magnetic susceptibility $\chi$ measurements for temperatures $T \leq 1000$~K we observe that for Li$_2$RuO$_3$, $\chi_{||} > \chi_{\perp}$.  Additionally, we observe a first-order high temperature coupled magneto-structural transition which seems to occur in two steps.  This is most evident in the $\chi_{\perp}$ data.  The higher temperature transition has an onset as high as $T > 570$~K and a mid-point around $T \approx 561$~K as seen by the peak in $d\chi_{\perp}/dT$ measured while warming up to $1000$~K\@.  This high temperature transition is hysteretic with a thermal hysteresis of $6$~K indicating its first-order nature.  The lower temperature transition in $\chi_{\perp}$ occurs at $T \approx 538$~K, is very sharp, is accompanied by an abrupt fall in $\chi$, and with almost no thermal hysteresis.  These observations suggest that the higher temperature, hysteretic transition is the structural dimerization transition while the lower temperature transition where we observe a sharp fall in $\chi$ is the magnetic transition involving Ru-Ru singlet formation.  Thus for Li$_2$RuO$_3$ the two transitions most likely occur at slightly different temperatures with the structural dimerization transition occuring first and triggering the magnetic Ru-Ru singlet formation.  The onset temperature of $570$~K is much higher than previously observed ($\approx 540$~K) and indicates the high quality of the samples.  

Just a $5\%$ substitution of Na for Li leads to interesting magnetic and structural changes.  The high temperature $\chi(T)$ data show that the magnetic anisotropy is reversed compared to Li$_2$RuO$_3$ with $\chi_{\perp} > \chi_{||}$ for (Li$_{0.95}$Na$_{0.05}$)$ _{2} $RuO$_3$.  The arrangement of Ru-Ru dimers on the honeycomb lattice also changes.  For Li$_2$RuO$_3$ Rietveld refinements of room temperature powder X-ray data reveal that one ($d_2$) out of the three inequivalent Ru-Ru bonds on the honeycomb lattice is shortened compared to the other two which are almost equal to each other as can be seen in Table~\ref{Table-lattice-parameters}.  For (Li$_{0.95}$Na$_{0.05}$)$ _{2} $RuO$_3$ we find that two ($d_2$ and $d_3$) out of the three Ru-Ru bonds are smaller and almost equal while the third is much larger.  The armchair arrangement of the dimers in Li$_2$RuO$_3$ is consistent with previous reports \cite{Miura2007, Jackli2008}.  The dimer arrangement in (Li$_{0.95}$Na$_{0.05}$)$ _{2} $RuO$_3$ can be viewed as two inter-penetrating armchairs formed on the $d_2$ and $d_3$ bonds, respectively.  This suggests a possible change in the orbital ordering pattern for the Na substituted sample.

\paragraph{Acknowledgments.--} We thank the X-ray facility at IISER Mohali for powder XRD measurements.  YS acknowledges DST, India for support through Ramanujan Grant \#SR/S2/RJN-76/2010 and through DST grant \#SB/S2/CMP-001/2013.  KM acknowledges UGC India for a fellowship.


\begin{references}

\bibitem{Jackeli2009} G. Jackeli, and G. Khaliullin, Phys. Rev. Lett. {\bf 102}, 017205 (2009).

\bibitem{Shitade2009} A. Shitade, H. Katsura, J. Kunes, X.-L. Qi, S.-C. Zhang, and N. Nagaosa, Phys. Rev. Lett. {\bf 102}, 256403 (2009).

\bibitem{Pesin2010} D. A. Pesin and Leon Balents, Nature Phys. {\bf 6}, 376 (2010).

\bibitem{Wan2011} X. Wan, A. M. Turner, A. Vishwanath, S. Y. Savrasov, Phys. Rev. B {\bf 83}, 205101 (2011).

\bibitem{Kitaev2006} A. Kitaev, Ann. Phys. {\bf 321}, 2 (2006). 

\bibitem{Chaloupka2010} J. Chaloupka,  G. Jackeli, and G. Khaliullin, Phys. Rev. Lett.   {\bf 105}, 027204 (2010).

\bibitem{Singh2010} Y. Singh, and P. Gegenwart, Phys. Rev. B { \bf 82}, 064412 (2010).

\bibitem{Singh2012} Y. Singh, S. Manni, J. Reuther, T. Berlijn, R. Thomale, W. Ku, S. Trebst, and P. Gegenwart. Phys. Rev. Lett. {\bf 108}, 127203 (2012).

\bibitem{Chun2015} S. H. Chun, J-W Kim, J. Kim, H. Zheng, C. Stoumpos, C. Malliakas, J. F. Mitchell, Kavita Mehlawat, Yogesh Singh, Y. Choi, T. Gog, A. Al-Zein, M. Moretti Sala, M. Krisch, J. Chaloupka, G. Jackeli, G. Khaliullin, and B. J. Kim, Nature Physics {\bf 11}, 3322 (2015).

\bibitem{Banerjee2016} A. Banerjee,	C. A. Bridges,	J.-Q. Yan,	A. A. Aczel,	L. Li,	M. B. Stone,	G. E. Granroth,	M. D. Lumsden,	Y. Yiu,	J. Knolle,	S. Bhattacharjee,	D. L. Kovrizhin,	R. Moessner,	D. A. Tennant,	D. G. Mandrus, and S. E. Nagler, Nature Materials (2016) doi:10.1038/nmat4604.

\bibitem{Gupta2014} S. Nath Gupta,  P. V. Sriluckshmy, K. Mehlawat, A. Balodhi, D. K. Mishra, D.V.S. Muthu, S. R. Hassan, Y. Singh, T. V. Ramakrishnan, and A. K. Sood, arXiv:1408.2239 (2014).

\bibitem{Sandilands2015} L. J. Sandilands, Y. Tian, K. W. Plumb, Young-June Kim, and K. S. Burch, Phys. Rev. Lett. {\bf 114}, 147201 (2015).

\bibitem{Mehlawat2015} Kavita Mehlawat, G. Sharma,  and Yogesh Singh, Phys. Rev. B {\bf 92}, 134412 (2015).

\bibitem{Lei2014} Hechang Lei, Wei-Guo Yin, Zhicheng Zhong, and Hideo Hosono, Phys. Rev. B {\bf 89}, 020409(R) (2014).

\bibitem{Mogare2004} K. M. Mogare, K. Friese, W. Klein, and M. Jansen, Z. Anorg. Allg. Chem. {\bf 630}, 547 (2004).

\bibitem{Wang2014} J. C. Wang, J. Terzic, T. F. Qi, F. Ye, S. J. Yuan, S. Aswartham, S. V. Streltsov, D. I. Khomskii, R. K. Kaul, and G. Cao, Phys. Rev. B {\bf 90}, 161110 (2014).

\bibitem{Dulac1970} J. F. Dulac, C. R. Acad. Sci. Paris, Ser. B {\bf 270}, 223 (1970).

\bibitem{James1988} A. C. W. P. James and J. B. Goodenough: J. Solid State Chem. {\bf 74}, 287 (1988).

\bibitem{Felner2002} I. Felner and I. M. Bradaric, Physica B {\bf 311}, 195 (2002).

\bibitem{Miura2007} Y. Miura, Y. Yasui, M. Sato, N. Igawa, and Kazuhisa Kakurai, J. Phys. Soc. Jpn. {\bf 76}, 033705 (2007).

\bibitem{Miura2009} Y. Miura,  M. Sato, Y. Yamakawa, T. Habaguchi, and Y. Ono, J. Phys. Soc. Jpn. {\bf 78}, 094706 (2009).  

\bibitem{Jackeli2008} G. Jackeli, and D. I. Khomskii, Phys. Rev. Rev. Lett. {\bf 100}, 147203 (2008).

\bibitem{Kimber2014} S. A. J. Kimber, I. I. Mazin, Juan Shen,  H. O. Jeschke, S. V. Streltsov, D. N. Argyriou, R. Valenti, and D. I. Khomskii, Phys. Rev. B, {\bf 89}, 081408(R) (2014).

\bibitem{Park2016} J. Park, T-Y. Tan, D. T. Adroja, A. Daoud-Aladine, S. Choi, D-Y. Cho, S-H. Lee, J. Kim, H. Sim, T. Morioka, H. Nojiri, V. V. Krishnamurthy, P. Manuel, M. R. Lees, S.V. Streltsov, D.I. Khomskii, J-G. Park, arxiv:1604.04019 (2016).

\bibitem{Pchelkina2015} Z. V. Pchelkina, A. L. Pitman, A. Moewes, E. Z. Kurmaev, Teck-Yee Tan, D. C. Peets, Je-Geun Park, and S. V. Streltsov, Phys. Rev. B {\bf 91}, 115138(2015).

\bibitem{Segura2016} M.-P. Jimenez-Segura, A. Ikeda, S. Yonezawa, and Y. Maeno, Phys. Rev. B {\bf 93}, 075133(2016).


\end{references}
\end{document}